\RequirePackage{fix-cm}
\documentclass[twocolumn,epjc3]{svjour3}  
\smartqed  
\RequirePackage{graphicx}
\usepackage{graphicx}
\usepackage{epstopdf}
\usepackage{mathrsfs}
\usepackage{amssymb}
\usepackage{verbatim}
\usepackage{color}
\usepackage{multirow}
\usepackage{cite}
\usepackage{hyperref}
\hypersetup{colorlinks,bookmarksopen,bookmarksnumbered,citecolor=blue,
linkcolor=black,pdfstartview=FitH,urlcolor=blue}
\journalname{Eur. Phys. J. C}
\begin{document}

\title{X-ray lines and self-interacting dark matter
}


\author{Yann Mambrini
\thanksref{e1,addr1}
        \and
        Takashi Toma
        \thanksref{e2,addr1} 
}

\thankstext{e1}{e-mail: yann.mambrini@th.u-psud.fr}
\thankstext{e2}{e-mail: takashi.toma@th.u-psud.fr}


\institute{Laboratoire de Physique Th\'eorique, CNRS - UMR 8627,\\  
Universit\'e de Paris-Sud 11, F-91405 Orsay Cedex, France \label{addr1}
}

\date{Received: date / Accepted: date}

\maketitle

\begin{abstract}
We study the correlation between a mono-chromatic
 signal from annihilating dark matter and its self-interacting 
cross section. We apply our argument to
 a complex scalar dark sector, where the pseudo-scalar plays the role
of a warm dark matter candidate while the scalar mediates its
 interaction with the Standard Model.
We combine the recent observation of the cluster Abell 3827 for
 self-interacting dark matter and 
 the constraints on the annihilation cross section for monochromatic
 X-ray lines. 
 We also confront our model to a set of recent experimental analyses and 
 find that such an extension can naturally produce
 a monochromatic keV signal
corresponding to recent observations of Perseus or Andromeda
 while in the meantime predicts self-interacting cross 
section of the order of  $\sigma / m \simeq 0.1-1~\mathrm{cm^2/g}$ 
 as recently claimed in the observation of the cluster Abell 3827.
 We also propose a way to distinguish such models by future direct
 detection techniques.

\end{abstract}

\section{Introduction}
\label{intro}

\noindent
Dark matter is inferred to exist, through its gravitational interactions
with visible matter, within and between
galaxies~\cite{Hinshaw:2012aka,Ade:2013zuv,Iocco:2015xga}. Even if the
PLANCK 
satellite~\cite{Ade:2013zuv} confirmed that about 85 \% of the total 
amount of the matter is dark, the community still lacks a clear evidence
of its nature through a direct or
indirect signal. 
Indeed, the last results of XENON100~\cite{Aprile:2012nq},
LUX~\cite{Akerib:2013tjd} and FERMI observation of 
the galactic center~\cite{FERMIGC} or dwarf galaxies~\cite{FERMIdwarfs} 
impose very strong constraints on the mass of a  weakly interacting
massive particle,
(if one excludes the  3$\sigma$ galactic center excess consistent with
the range of dark matter identified in the FERMI-LAT
data~\cite{Daylan:2014rsa}), 
questioning the WIMP paradigm. Little is known about the
mass
and coupling of dark matter, and even the 
``WIMP miracle'' is questionable~\cite{Feng:2008ya} if one
introduces a hidden mediator sector ``$X$'', with its mass $m_X$ and 
coupling $g_X$ respecting $m_X/g_X \simeq m_\mathrm{wimp}/g_{EW}$ where
$m_\mathrm{wimp}$ is the WIMP mass and $g_{EW}$ is the electroweak gauge
coupling constant. Much lighter and warmer 
candidates are then allowed, and can justify the lack of GeV signal in
direct and indirect detection experiments, while
explaining in the meantime recent claims at the keV scale~\cite{Boehm:2003bt}. 

A possible smoking gun signature of the interaction of dark matter in our
galaxy or in larger structure would be 
the observation of a monochromatic signal (photon, neutrino or positron) 
generated by the annihilation or the decay of the candidate. In 2012,
several studies claimed for the
observation of
a $135~\mathrm{GeV}$ monochromatic photon-line produced near the center of our
Milky Way~\cite{Wenigerline}. Phenomenological models describing
the
possibility of generating such a line then appeared in the
literature~\cite{Dudas:2012pb,otherwenigerline}.
More recently, the presence of a seemingly unexplained  X-ray line observed 
by the XMM-Newton observatory in galaxies and
galaxy clusters \cite{Bulbulline} 
increased the interest for  
annihilating~\cite{Dudas:2014ixa,annihilatingbulbulline} or
decaying~\cite{decayingbulbulline} light dark matter scenarios. 
Excited dark matter~\cite{excitingbulbulline} or axion-like
candidates~\cite{axionbulbulline} were also proposed as alternative 
interpretations.\footnote{To keep the analysis as fair as possible, it is important to 
underline that there are still on-going debates on
the possibility of explaining the X-ray line excess with thermal atomic
transition~\cite{Jeltema:2014qfa}.}

On the other hand, if the X-ray line excess discussed above is interpreted as a dark matter signal,
a same excess should be observed from the other galaxies such as the Milky
Way, M31 and dwarf spheroidal galaxies in addition to
the Perseus and Centaurus clusters. 
However such a signal has not yet been observed in Milky
Way~\cite{Riemer-Sorensen:2014yda}, M31~\cite{Horiuchi:2013noa}, stacked 
galaxies~\cite{Anderson:2014tza} and stacked dwarf galaxies~\cite{Malyshev:2014xqa}. 
For completeness, keeping open all possible interpretation of the 3.5 keV
line signal, we will present the result of both analysis (signal or constraint) in
every scenarios we studied in this work.

In parallel, recently the authors of~\cite{Massey:2015dkw} claimed that
the observations
of one (particularly well constrained) galaxy in the cluster Abell 3827
revealed a surprising $\simeq 1.62$ kpc offset between its dark
matter and stars. They affirm that such an offset is consistent with
theoretical predictions from the models of self-interacting dark matter,
implying a lower bound of the self-interacting cross section divided by
the dark matter mass $\sigma / m \gtrsim
10^{-4}~\mathrm{cm^2/g}$. In the meantime, another
group~\cite{Kahlhoefer:2015vua} with a different kinematical
analysis for the very same galaxy obtained the value $\sigma / m \gtrsim
1.5~\mathrm{cm^2/g}$ in  the case of contact interaction corresponding
to the exchange of a massive mediator in opposition to long range
interaction which can arise for example 
from a massless mediator~\cite{Feng:2009mn}. Entering into 
the debate of the exact value deduced from the observations is far
beyond the scope of our work. However, one has to admit 
that any evidence for dark matter self-interaction would have strong
implications for particle physics, as it would severely constrain or
even rule out popular candidates such as supersymmetric
neutralino/gravitino, axions, or any Higgs, 
$Z$, $Z'$ portal WIMP--like
candidates. The main reason
is that, within the sensitivity of present measurements, the
observation of a self-interaction would imply the ratio 
$\sigma / m \simeq (10^{-5}-2) ~ \mathrm{cm^{2} g^{-1}} \simeq
(0.05-9000)~\mathrm{GeV^{-3}}$, which is much larger than any typical WIMP values 
$\sigma_\mathrm{wimp}/m_\mathrm{wimp} \simeq 10^{-11}~ \mathrm{GeV^{-3}}$.

In this work, we show that it is possible, in a minimal
framework, to relate naturally the (smoking gun) monochromatic signal 
generated by the annihilation
of a pseudo-scalar particle to its self-interaction process. 
As a consequence, any signal or constraint derived
by the (non-)observation
of self-annihilation (coming for instance from the ``Bullet Cluster''
(1E 0657-56) which is typically of the order 
 of $\sigma / m \lesssim 1 ~\mathrm{cm^2 /
 g}$~\cite{Markevitch:2003at,Randall:2007ph}) 
 induces direct limits on the monochromatic signature. 
We begin our study by combining the constraints from different
experimental analyses, 
  before applying our results to the recent 3.5 keV line claims
\cite{Bulbulline}. We show that the observation of such a signal
implies naturally a relatively strong self-interacting process
compatible
with the limits on $\sigma/m$ obtained recently \cite{Massey:2015dkw,Kahlhoefer:2015vua}.
We would like to insist that beyond the 3.5 keV signal consideration
(one does not need to agree with the dark matter interpretation of the
line or the self-interacting dark matter observations) the aim of our work is more
general. We show the correlation which exists 
between an indirect detection signal and a self-interacting process
once one builds an explicit microscopic model,
 with a dynamical
symmetry breaking, which are not necessary present if one takes a
pure effective approach\footnote{Interestingly, the authors in~\cite{Boddy:2014qxa}
 addressed a similar issue in the case of exciting dark
matter, and long range interaction. Our framework being annihilating
dark matter  and 
contact interaction, our model, discussions, results and prospect are
completely different.}.

The paper is organized as follows. After a short description of the
model under consideration in section II, we compute and analyze the self-interaction
process combined with 
the monochromatic constraints and signal extracted from a set of
different experimental collaborations in section III. Section IV is
devoted to the
discussion and signatures in terms of indirect and direct detection
prospects in more general cases.  We draw our conclusions in section V,
while Appendix contains alternative scenarios with a fermionic dark
matter.

 \section{The framework}
 \subsection{Minimal model} 

\noindent
In this section, we describe the model of a pseudo-scalar dark matter.
The reader interested in alternative scenarios can find in 
Appendix the formulae in the case of a fermionic dark matter. The
model was originally built with success to interpret 
the recent monochromatic signal observed in different clusters of
galaxies~\cite{Dudas:2014ixa}. In this model, 
a scalar, or pseudo-scalar particle 
is {\it by definition} a self-interacting particle. The Higgs boson,
unique observed spin 0 particle until now, is a self-interacting particle through
its quartic coupling. Several other self-interacting candidates have
been proposed in the literature, but usually these were spin 1/2
particles. However, in this case, it becomes necessary to invoke specific processes (like
Sommerfeld enhancement, or strong interaction)
to compensate the dimensionality of the 4-fermion couplings.
In the case of a scalar or pseudo-scalar dark matter $\phi$, the self
interaction term $\frac{\lambda}{4} |\phi|^4$ is always allowed by
a global $U(1)$ invariance and induces then necessarily self-interacting
processes. Moreover, in the framework of spontaneous symmetry breaking, 
a strong correlation exists between the vacuum expectation value ($vev$)
of $\phi$, its mass and the quartic coupling $\lambda$, rendering the
construction very predictive. 

The general renormalizable potential for a scalar complex field $|\Phi
|^2$ respecting a global $U(1)$ symmetry is\footnote{We neglected
throughout our study the possible Higgs-mixing 
as recent analysis on the invisible width of the Higgs impose stringent
constraints on such mixings~\cite{Robens:2015gla}.} 

\begin{equation}
\mathcal{V}_{\Phi}=-\mu^2|\Phi|^2+\frac{\lambda}{4}|\Phi|^4,
\end{equation}
where $\mu^2$ is the bare mass of $\Phi$ and $\lambda$
is the quartic coupling of $\Phi$.

\noindent
After a spontaneous breaking of the symmetry, it is
straightforward to re-express the potential as a function
of the fundamental components of $\Phi = v + \frac{s + i a}{\sqrt{2}}$ 
with $v= \langle \Phi \rangle = \sqrt{\frac{2}{\lambda}} \mu$. 
Absorbing the unphysical constants, we obtain

\begin{eqnarray}
\mathcal{V}_{\Phi}&=& \frac{m_s^2}{2} s^2 + \frac{\sqrt{\lambda}}{2
\sqrt{2}} m_s s^3
+ \frac{\sqrt{\lambda}}{2 \sqrt{2}} m_s a^2 s 
\nonumber\\
&&
+ \frac{\lambda}{16} s^4 + \frac{\lambda}{16} a^4
+ \frac{\lambda}{8} a^2 s^2,
\label{Eq:potential}
\end{eqnarray}

\noindent
with the scalar mass $m_s=\sqrt{2} \mu = \sqrt{\lambda} v$. 
It is important to notice that
if our $U(1)$ symmetry was exact (prior to develop a $vev$), the
pseudo-scalar dark matter mass $m_a$ would remain massless to all orders in
perturbation theory. In what follows, we will assume
that the $U(1)$ symmetry is broken by non-perturbative effects down to a
discrete $Z_N$ symmetry. 
It is actually standard in string theory that all symmetries are gauged
symmetries in the UV.\footnote{See~\cite{Dudas:2014bca} for a concrete
example in the same framework
where it has been shown that, in the meantime, a hierarchy $m_a \ll m_s$
is generated by the mechanism.} 
Thus a non-zero dark matter mass $m_a$ being much lighter than $m_s$ is
expected.

\subsection{The self-interaction process}

\begin{figure}
\begin{center}
\includegraphics[scale=0.8]{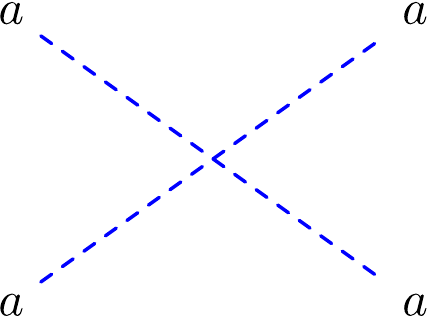}
\hspace{0.5cm}
\includegraphics[scale=0.8]{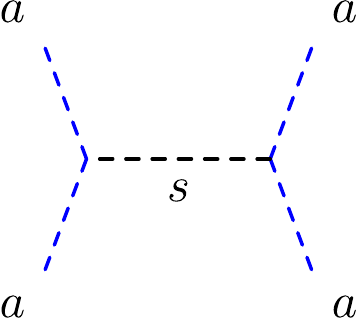}\\
\vspace{0.5cm}
\includegraphics[scale=0.8]{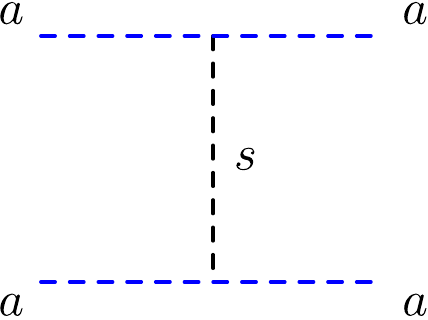}
\hspace{0.5cm}
\includegraphics[scale=0.8]{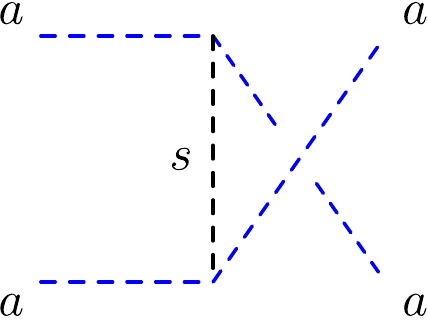}
\caption{Feynman diagrams for dark matter self-interacting cross
 section.}
\label{Fig:aa}
\end{center}
\end{figure}

 \noindent
 In our model, we have four diagrams contributing to the
 self-interacting cross section as depicted in Fig.~\ref{Fig:aa}. 
 Once the scalar part of $\Phi$ develops a $vev$ it becomes possible to
 re-express the total cross section as
 
\begin{eqnarray}
\frac{\sigma_{aa}}{m_a}&=&\frac{\lambda^2 m_a}{32 \pi m_s^4 \left(1-4
\frac{m_a^2}{m_s^2}\right)^2}
\nonumber
\\
&\simeq& \frac{\lambda^2 m_a}{32 \pi m_s^4},~~~~~~~ (m_s \gg m_a).
\label{Eq:sigmaaa}
\end{eqnarray}

\noindent
It is interesting to note that the cross section is of the form
$\sigma_{aa} \propto  m_a^2/m_s^4$ and then null for $m_a=0$,
whereas if one takes into account only the quartic vertex $aaaa$, it should naively
be proportional to $1/m_a^2$ and could potentially diverge. 
The mechanism canceling the divergences is in fact similar to the Higgs
contribution occurring in the $WW$
scattering in the Standard Model. 
This can be easily understood  as $m_a$ can be considered as the
pseudo-goldstone boson generated by the breaking of the global $U(1)$ symmetry.  
This fundamental feature {\it would not} have been observed in the
framework of an effective approach if one introduces a dimensional coupling
of the form $\tilde \mu s aa $, $\tilde \mu$ being a
free mass parameter. It is thus the dynamical structure of the construction
which defines precisely its self--coupling constants. 
Another interesting point is that for a MeV scale mediator $s$, one does not
need to invoke very large 
values of $\lambda$ to obtain self-interacting cross section compatible
with recent analysis. For instance, in the case of $m_a = 3 $ keV 
and $m_s=1$ MeV, 
one obtains $\sigma_{aa}/m_a \simeq 7 \lambda^2 ~\mathrm{cm^2 /g}$ which is
in the order of the measured limit ($\sigma/m\lesssim1~\mathrm{cm^2/g}$)
for a reasonable value of $\lambda \simeq 1$, much below the
perturbativity limit, without invoking velocity enhancement.

\subsection{Monochromatic photon}

\noindent
Concerning the coupling to the photons, we consider the coupling which
can be written as 

\begin{equation}
\mathcal{L}_{s\gamma \gamma}=\frac{s}{\Lambda}F_{\mu\nu}F^{\mu\nu},
\label{eq:dim5}
\end{equation}

\noindent
with $F_{\mu \nu} = \partial_\mu A_\nu - \partial_\nu A_\mu$ 
being the electromagnetic field strength.
The scale $\Lambda$  can be interpreted in a UV
completion since it can be determined by a set of new heavy 
charged particles running in triangular loops. 
The mass scale of new charged particles is assumed to be heavier than
$300$ GeV to respect the LEP constraint, depending
on the number of charged particles. Several experiments restrict
 $\Lambda$ from the Horizontal Branch
(HB) stars processes~\cite{Masso:1995tw,Raffelt:1987yu,Cadamuro:2012rm}
to the LEP~\cite{Kleban:2005rj} or beam dump experiment 
constraints~\cite{Riordan:1987aw,Bjorken:1988as}. We will 
review them in detail in the next section, but roughly
speaking, the coupling of a scalar to photons is extremely suppressed
($\Lambda\gtrsim10^{10}~\mathrm{GeV}$) for $m_s \lesssim 300$ keV, largely
 due to the HB limits. 
For $m_s \gtrsim 300$ keV, a window opens, allowing values of
$\Lambda$ as low as 10 GeV.  In a UV complete
model, such low values of $\Lambda$ can be understood if the number of
fermions running in the loop is relatively important 
(of the order of 10). 

\begin{figure}
\begin{center}
\includegraphics[scale=0.8]{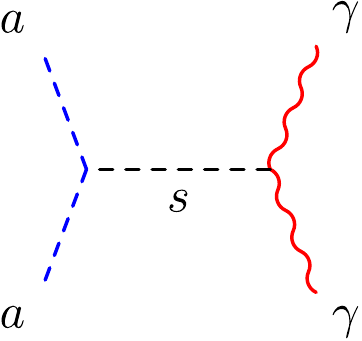}
\hspace{0.5cm}
\includegraphics[scale=0.8]{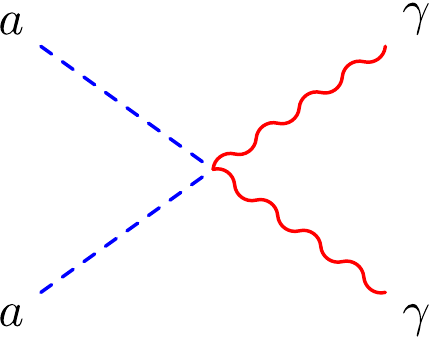}
\caption{Feynman diagrams for dark matter annihilation
 into two
 photons. The second diagram can be generated by higher dimensional
 operators (see the text for details).}
\label{Fig:gg}
\end{center}
\end{figure}

The presence of $s A_\mu A_\nu$ coupling generates naturally the
production of monochromatic photons from the s-channel annihilation
 of the dark matter candidate $a$ as depicted in the left of Fig.~\ref{Fig:gg}.
The annihilation cross section for $aa\to\gamma\gamma$ is given
by~\cite{Dudas:2014ixa}

\begin{equation}
\sigma{v}_{\gamma\gamma}=\frac{\lambda
m_a^2m_s^2}{\pi\Lambda^2(m_s^2-4m_a^2)^2}.
\label{eq:sigmav_gg}
\end{equation}

\noindent
For $m_a\ll m_s$, the above cross sections Eq.~(\ref{Eq:sigmaaa}) and
(\ref{eq:sigmav_gg}) can be simplified to be 
\begin{equation}
\frac{\sigma_{aa}}{m_a}\approx\frac{\lambda^2 m_a}{32\pi m_s^4},\qquad
\sigma{v}_{\gamma\gamma}\approx\frac{\lambda m_a^2}{\pi\Lambda^2m_s^2}.
\end{equation}

 \noindent
By eliminating $\lambda$ in both expressions, it becomes possible for
each energy $E_\gamma$ being equivalent to the dark matter mass $m_a$
since dark matter is almost at rest, to express $\sigma v_{\gamma \gamma} (E_\gamma)$
{\it uniquely} as a function of $\Lambda$ and $\sigma_{aa}/m_a$,

 \begin{eqnarray}
 \sigma v_{\gamma \gamma} &=& \frac{4 \sqrt{2} E_\gamma^{3/2}}{\Lambda^2
 \sqrt{\pi}} \sqrt{\frac{\sigma_{aa}}{m_a}}\nonumber
 \\
 &\simeq&
 1.3 \times 10^{-33} \left( \frac{100~\mathrm{TeV}}{\Lambda} \right)^2
 \left(\frac{E_\gamma}{3~\mathrm{keV}}\right)^{3/2}\nonumber\\
  &&\times
 \sqrt{\frac{\sigma_{aa}/m_a}{1~\mathrm{cm^2 /g}}} ~~ \mathrm{cm^3/s} .
\label{Eq:sigsigv}
 \end{eqnarray}

 This is one of the main results of our work. It is indeed surprising
 that asking for a reasonable value for the self-interacting cross section 
 of the order of $1~ \mathrm{cm^2/g}$, one obtains naturally the
 annihilation cross section of the order of $10^{-33}~\mathrm{cm^3
 s^{-1}}$ for a monochromatic keV signal, which corresponds exactly to
 the magnitude of the signals observed by XMM Newton~\cite{Bulbulline}
 in the Perseus cluster.\footnote{It is also interesting to note the
 possibility to obtain in the meantime the suitable relic abundance of
 dark matter without affecting $N_\mathrm{eff}$ if one adds a coupling to the
 neutrino sector as was shown in \cite{Dudas:2014ixa}.}
  On the other hand,
 strong limits obtained from the non-observation of a monochromatic line
  by observatories  such as HEAO-1 INTEGRAL, COMPTEL, EGRET and FERMI
 restrict severely the lower bound on the scale 
 $\Lambda$ in the rest of the parameter space, as we will analyse in the
 following section.

\subsection{A remark on higher-dimensional operator analysis}
\noindent
Building a complete ultraviolet model  is far beyond the scope of this
work, but we can give
some hints for further developments.
Indeed, even if the Lagrangian Eq.~(\ref{eq:dim5}) breaks $explicitly$
the $U(1)$ symmetry,
we can have a look to higher dimensional operators 
which can generate such a term after the breaking of the $U(1)$
symmetry. 
The simplest dimension $6$ operator can be written as 
\begin{equation}
\mathcal{L}_{\Phi\gamma\gamma}=\frac{|\Phi|^2}{\tilde{\Lambda}^2}F_{\mu\nu}F^{\mu\nu},
\label{Eq:dim6langrangian}
\end{equation}
with $\tilde{\Lambda}$ being a different cut-off scale from $\Lambda$
introduced in Eq.~(\ref{eq:dim5}). 
After the symmetry breaking, one  obtains the interaction terms
\begin{equation}
\mathcal{L}_{\Phi\gamma\gamma}\supset
\left(\sqrt{\frac{2}{\lambda}}\frac{m_ss}{\tilde{\Lambda}^2}+\frac{a^2}{2\tilde{\Lambda}^2}\right)
F_{\mu\nu}F^{\mu\nu}. 
\label{eq:dim6}
\end{equation}
One can then deduce from Eq.~(\ref{eq:dim6}) the relation between
$\Lambda$ and $\tilde{\Lambda}$: 
$\Lambda=\sqrt{\frac{\lambda}{2}}\frac{\tilde{\Lambda}^2}{m_s}$.
The effective model built from the Lagrangian 
generates the second term in Eq.~(\ref{eq:dim6}). This contact
interaction contributes also to the annihilation cross section
$aa\to\gamma\gamma$. 
Including this new contribution, the total cross-section is then given 
by
\begin{equation}
\sigma{v}_{\gamma\gamma}=\frac{32m_a^6}{\pi\tilde{\Lambda}^4\left(4m_a^2-m_s^2\right)^2}.
\end{equation}
However, in the rest of our work we will continue to consider 
the  dimension-5 coupling approach
$\frac{s}{\Lambda} F^{\mu \nu} F_{\mu \nu}$ 
because a complete dimension-6 operator analysis would require a
much more careful study of all the possible operators
involved in the processes.

 \section{The measurements}

  \subsection{Self-interacting dark matter}
 
 \noindent
 The status of the (non-)observation of self-interacting dark matter has
 become somewhat quite confusing recently, due
 to the release of (seemingly) contradictory results. Indeed, some
 authors of Ref.~\cite{Massey:2015dkw} using the new Hubble Space Telescope
 imaging,
  claimed to have
 observed that the dark matter halo of at least one of the central
 galaxies belonging to the cluster Abell 3827 is spatially offset from
 its stars. The offset, of the order of 1.62 kpc could be interpreted as
 an evidence of self-interacting dark matter with a ratio of 
 cross section over mass 
 of $\sigma / m \simeq 1.7 \times 10^{-4} ~\mathrm{cm^2 / g}$.\footnote{It
 is interesting to note that, in the meantime, the same authors derived
 recently before a stringent bound on the self-interacting 
 scenario by the observation of 72 clusters collisions of $\sigma / m
 \lesssim 0.47 ~\mathrm{cm^2 / g}$~\cite{Harvey:2015hha}.} In the meantime,
 using a different kinematical approach
 than~\cite{Massey:2015dkw}, the authors of~\cite{Kahlhoefer:2015vua}
 obtained a value  of $\sigma / m \simeq (1.5-3)~\mathrm{cm^2 /
 g}$, resulting in 
 tension with the upper bounds set by other astrophysical objects such
 as the ``Bullet Cluster'' (1E 0657-56) which are typically of the order
 of $\sigma / m \lesssim 1 ~\mathrm{cm^2 /
 g}$~\cite{Massey:2015dkw,Harvey:2015hha,Markevitch:2003at,Randall:2007ph,Rocha:2012jg,Peter:2012jh}. 
 The main difference between 
 the two analyses came from some approximations concerning 
 the evolution time, and the gravitational back-reaction of the halo on
 its stars during the separation process due the drag forces. The
 authors of Ref.~\cite{Kahlhoefer:2015vua} have already addressed this issue some
 time ago in~\cite{Kahlhoefer:2013dca}. They clearly distinguished 
 the contact interaction or $rare$ interaction (our case) from the
 long-range force (involving Sommerfeld enhancement) or $frequent$ 
 self-interactions through the position of the peak of the dark matter
 distribution compared to the position of the stars/galaxies after the
 interaction.
 
  In this work, we decided to
 take the two values proposed by the two groups as benchmark points, to
 show the correlation between an indirect signal 
 (monochromatic photon in our case)  and the self-interaction, once one
 has built an explicit microscopic model. 
 Some recent phenomenological constructions explaining these
 observations can be found in~\cite{Choi:2015bya} for a model with 
 a gauged $Z_3$ discrete symmetry; 
 a very nice interpretation in the framework of a Higgs portal
 (freeze-in mechanism) in~\cite{Campbell:2015fra} whereas other authors introduced a dark
 photon sector~\cite{D'Agnolo:2015koa} or a strong interacting
 sector~\cite{Bernal:2015bla}.

\subsection{Other experimental constraints}

\noindent
When a light (pseudo-)scalar interacts with photons, 
the helium burning period of HB stars is shortened 
due to non-standard energy loss since the light (pseudo-) scalar is
produced in the stellar interior by photons within thermal
distribution~\cite{Masso:1995tw,Raffelt:1987yu}. 
This effect gives a strong constraint on the coupling between the light
(pseudo-)scalar and photons for the (pseudo-)scalar mass lighter than $300~\mathrm{keV}$. 
The detailed analysis has been done in Ref.~\cite{Cadamuro:2012rm} and
we used their result in our study.\footnote{Note that the $aas$ coupling
in the 
scalar potential could also affect  the
HB bound since a pair of dark matter can be generated from
the off-shell produced scalar $s$. However, this off-shell 
contribution is expected to be much smaller than the dominant Primakoff
effect.
}

The interaction between the (pseudo-)scalar and photons is also
constrained by
mono-photon search at leptonic collider experiments.
Its signature is  $e^+e^-\to
\gamma+\mbox{\it missing energy}$. 
The collider bound has been compiled in Ref.~\cite{Masso:1995tw},
 taking into account the ASP (Anomalous Single
 Photon) experiment~\cite{Hearty:1989pq}. 
In addition, the improved LEP (Large Electron-Positron Collider)
limits based on the data of ALEPH, OPAL, L3 and DELPHI have
been published in Ref.~\cite{Kleban:2005rj}. 

\subsection{Relic abundance}
 
 \noindent
 The computation of the relic abundance of dark matter in our framework has already been
 studied in detail in \cite{Dudas:2014ixa}.
 We will not repeat the analysis in this work, but we recall its main
 point. Adding interactions to the neutrino sector
 through $s$ as a mediator can fulfill perfectly the relic abundance of
 dark matter measured by PLANCK
 while in the meantime generating 
 naturally a massive neutrino sector respecting the recent cosmological
 bounds on neutrino masses ($m_\nu \lesssim 1$ eV, see \cite{Lesgourgues:2012uu}
 for a review on the subject). The presence of a dark bath between the neutrino and 
 dark matter allows it in the parameter range of our work.
 However, keeping in mind this elegant possibility, our aim is to study
 the properties of self-interacting dark matter 
 at the present time and the correlation between 
 different observations in present large scale structures, independently
 on hypothesis concerning the thermal history
 generating the correct amount of dark matter abundance.

 \subsection{The 3.5 keV line signal}

\noindent
Recent claims for a detection of X-ray line observed in galaxies and
galaxy clusters like Perseus, by the XMM-Newton observatory~\cite{Bulbulline} 
increased the interest in light dark matter scenarios. Keeping in mind that
the status is still in debate (see the thermal atomic transition
interpretation
in \cite{Jeltema:2014qfa} for instance), it
is nevertheless interesting to apply our analysis in this concrete
example to check if
such a signal can be compatible with the limits derived from the recent
self-interaction measurements.

 The flux generated by the annihilation of dark matter in the Perseus
 cluster for instance can be computed 
 from the luminosity of the cluster $L$~\cite{Dudas:2014ixa}
 
\begin{eqnarray}
L &=& 
\int_{0}^{R_{Pe}} 4 \pi r^2 n^2_\mathrm{DM}(r) \langle \sigma v
\rangle_{\gamma \gamma}dr\nonumber\\
&=&
\int_{0}^{R_{Pe}} 4 \pi r^2 \left( \frac{\rho_\mathrm{DM}(r)}{m_a} \right)^2 \langle
\sigma v \rangle_{\gamma \gamma}dr,
\label{Eq:luminosity}
\end{eqnarray}

\noindent
with the Perseus radius $R_{Pe}$, the number density of dark matter
$n_\mathrm{DM}(r)$, the dark matter profile $\rho_\mathrm{DM}(r)$ and the thermally averaged cross
section $\langle\sigma{v}\rangle_{\gamma\gamma}$. At a first approximation, one
can consider a mean density of dark matter in the cluster as in
Ref.~\cite{Krall:2014dba}. The Perseus observation involved the mass 
of $M_{Pe}= 1.49 \times 10^{14} M_{\odot}$ in the region of $R_{Pe}=0.25$
Mpc at the distance of $D_{Pe}=78$ Mpc from the solar system. 
One can then estimate

\begin{eqnarray}
n_\mathrm{DM} &\simeq& \frac{1.49 \times 10^{14}M_{\odot}
}{m_a}\!\bigg/\frac{4 \pi R_{Pe}^3}{3}\nonumber\\
 &=& 1.9\times
10^{-37}~\mathrm{GeV^3}~~~(m_a=3.5~\mathrm{keV})\nonumber \\ 
&=& 2.5 \times 10^4~\mathrm{cm^{-3}} \ .\quad
\label{Eq:ndm}
\end{eqnarray}

\noindent
Combining Eq.~(\ref{Eq:luminosity}) and (\ref{Eq:ndm}), one can then compute
the luminosity in the Perseus cluster in the ``mean'' approximation
$\langle L\rangle$,

\begin{eqnarray}
\langle L\rangle &\simeq& 1.2 \times 10^{55}
\left( \frac{3.5~\mathrm{keV}}{m_a} \right)^2\nonumber\\
&&\times \left( \frac{\langle \sigma v \rangle_{\gamma \gamma}}{10^{-26}
 \mathrm{cm^3 s^{-1}}} \right)~\mathrm{photon/s}
\ .
\end{eqnarray}

\noindent
This estimation would be reasonable since dark matter in our
model does not have a cusp profile such as NFW or
Einasto, but a cored profile due to the large self-interacting 
cross section of dark matter. 
One can then deduce the flux $\phi_{\gamma \gamma} = L/(4 \pi D_{Pe}^2)$
that one should observe on earth

\begin{eqnarray}
\phi_{\gamma \gamma} &=& 1.6 \times 10^{-5} \left( \frac{3.5 ~
\mathrm{keV}}{m_a} \right)^2\nonumber\\ 
&&\times\left( \frac{\langle \sigma v \rangle_{\gamma \gamma}}{10^{-32}
\mathrm{cm^3 s^{-1}}} \right)~\mathrm{cm^{-2} s^{-1}}.
\end{eqnarray}

\noindent
According to the authors of Ref.~\cite{Bulbulline}, one can identify the
monochromatic signal arising from Andromeda galaxy (M31) 
 or Perseus cluster with the flux $\phi_{\gamma
 \gamma}=5.2_{-2.13}^{+3.70} \times 10^{-5}~\mathrm{cm^{-2}s^{-1}}$ 
 at 3.56 keV including the cluster core.\footnote{Another possibility
 would have been to 
 look at the Centaurus~\cite{Krall:2014dba} with $M_{Ce} = 6.3 \times
10^{13} M_{\odot}$ and a radius of $R_{Ce} = 0.17$ Mpc.} 
We will parametrize our uncertainty from the dark matter distribution in
the source by the classical ``astrophysical'' parameter $J_{astro} \ge
1$ 
with $J_{astro} = L/\langle L\rangle$, $L$ being the effective
luminosity for a steeper profile than the mean one we considered 
 above.

Finally, extending the analysis by taking into account also other
observations like M31, 
we will impose in our analysis a conservative annihilation cross section
which is required to reproduce the X-ray line estimated as

\begin{equation}
\langle \sigma v \rangle_{\gamma \gamma} \simeq \frac{1}{J_{astro}}(2
\times 10^{-33} - 8.5 \times 10^{-33}) \ \mathrm{cm^3 s^{-1}}. 
\label{Eq:sigv}
\end{equation}

 \section{The results}
 
 \subsection{Combining the line and self-interaction}

\begin{figure}
\begin{center}
\includegraphics[scale=0.65]{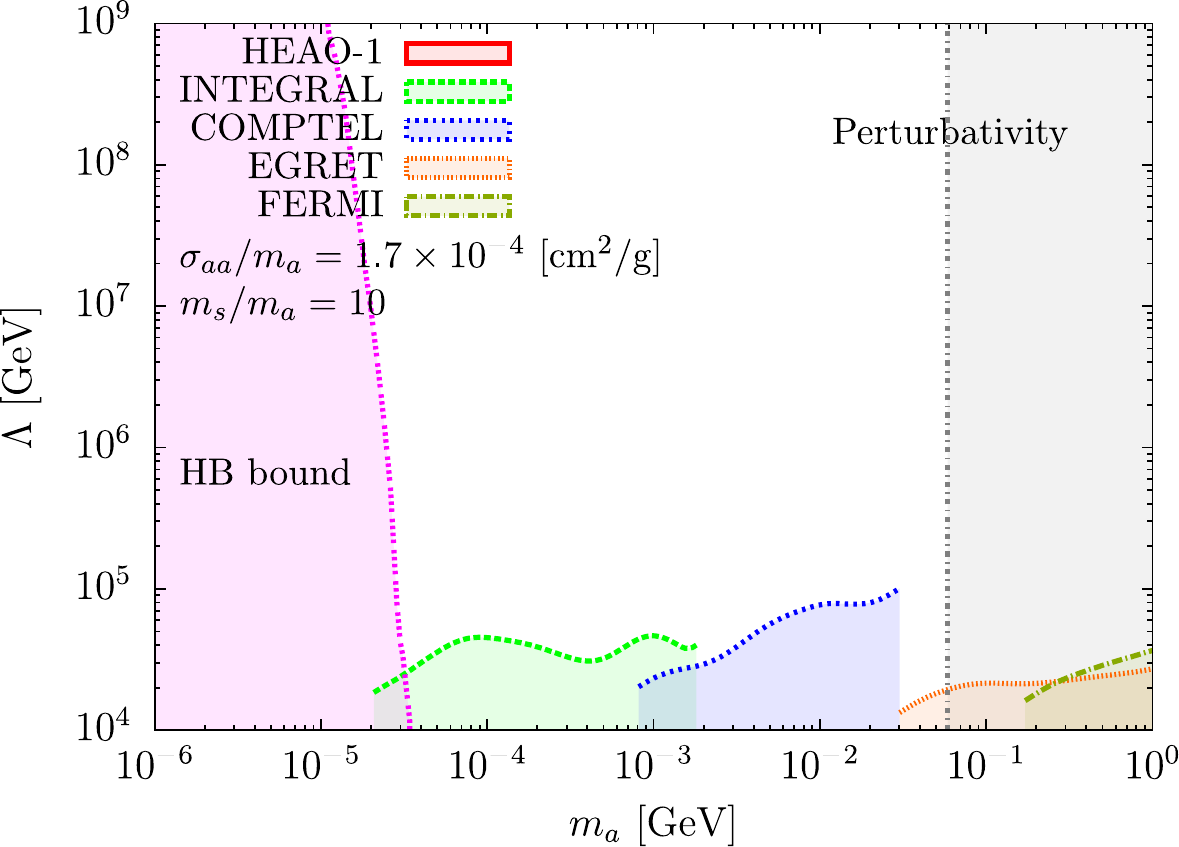}
\includegraphics[scale=0.65]{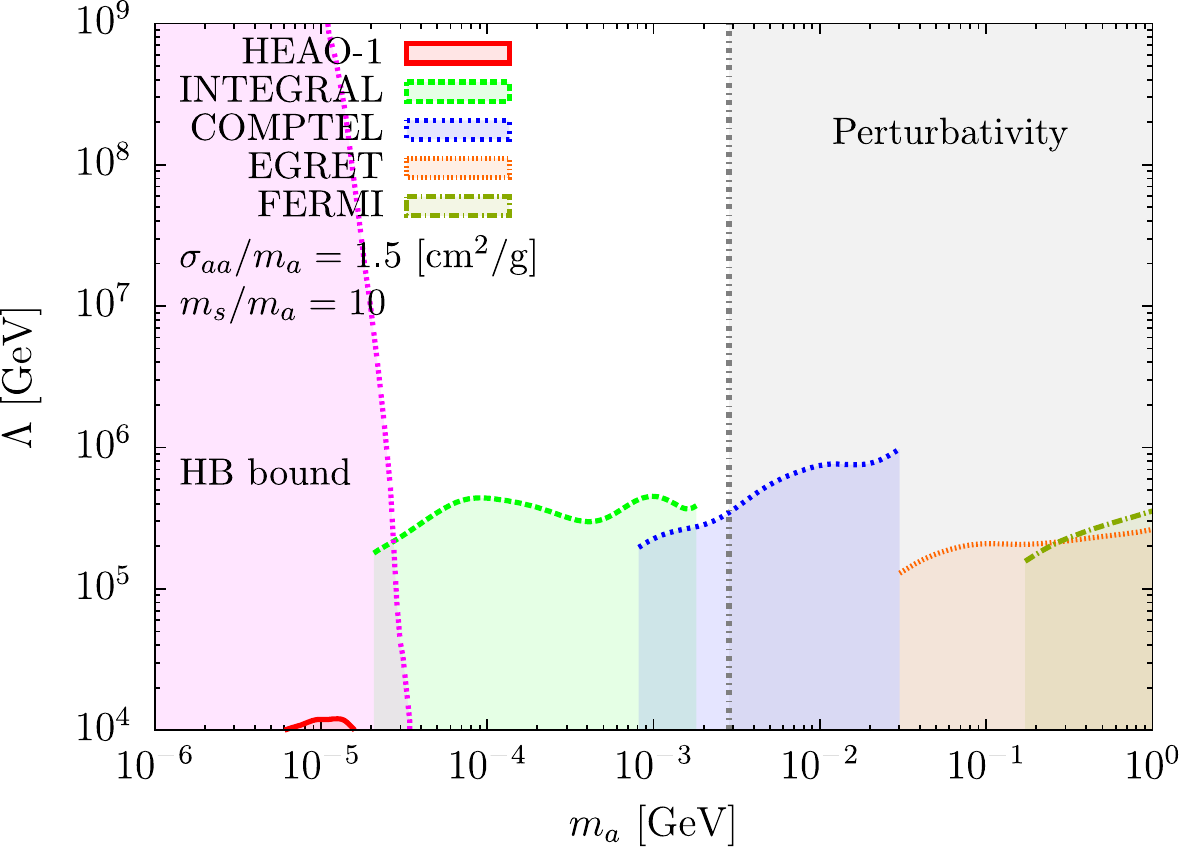}
\caption{
Limits on $\Lambda$ obtained from different observatories and
 satellites with the mass ratio fixed to $m_s/m_a=10$ where the white
 region is allowed and the colored region is excluded. 
The lower bounds of $\Lambda$ have been obtained from the
 data of satellites HEAO-1 (red), INTEGRAL (green), COMPTEL (blue),
 EGRET (brown), FERMI (dark-yellow). 
The HB bound (violet) and perturbativity bound (grey) for $\lambda$ are
 also shown. 
}
\label{Fig:idlimit}
\end{center}
\end{figure}

\begin{figure}
\begin{center}
\includegraphics[scale=0.65]{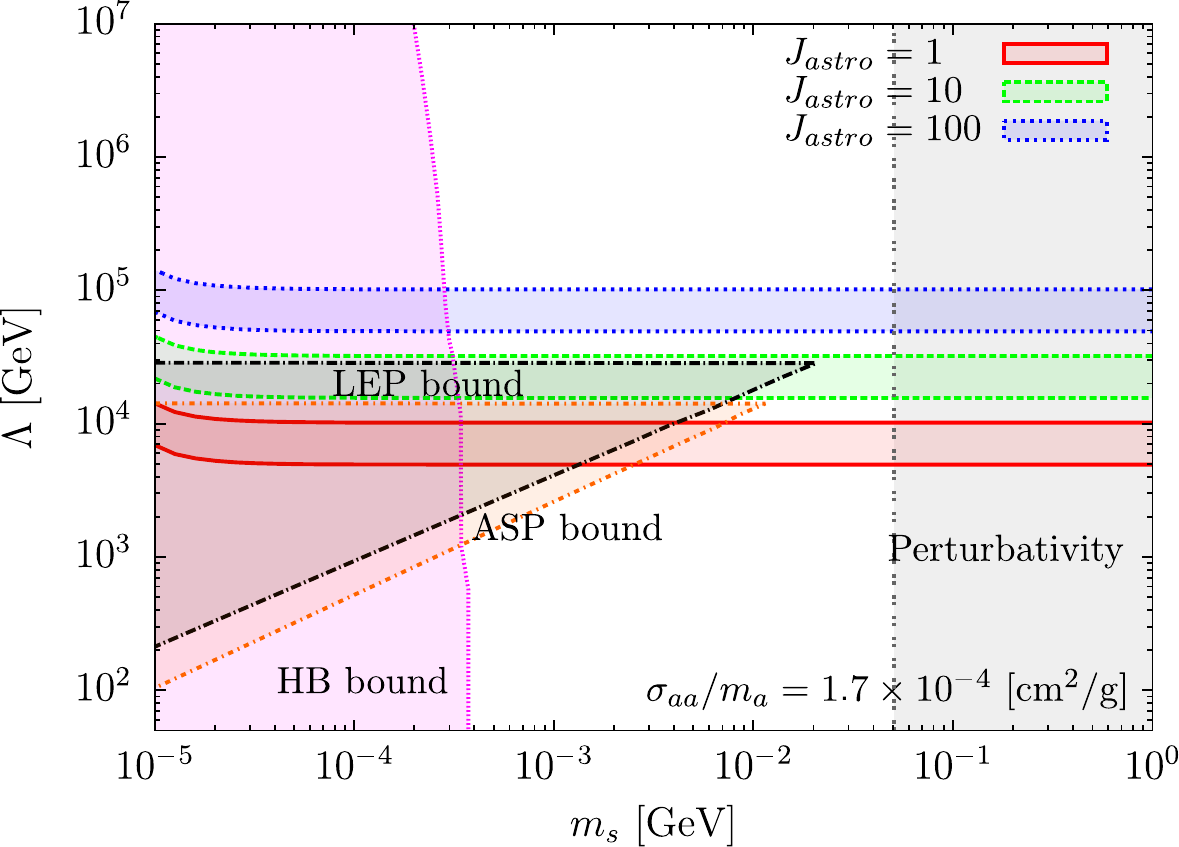}
\includegraphics[scale=0.65]{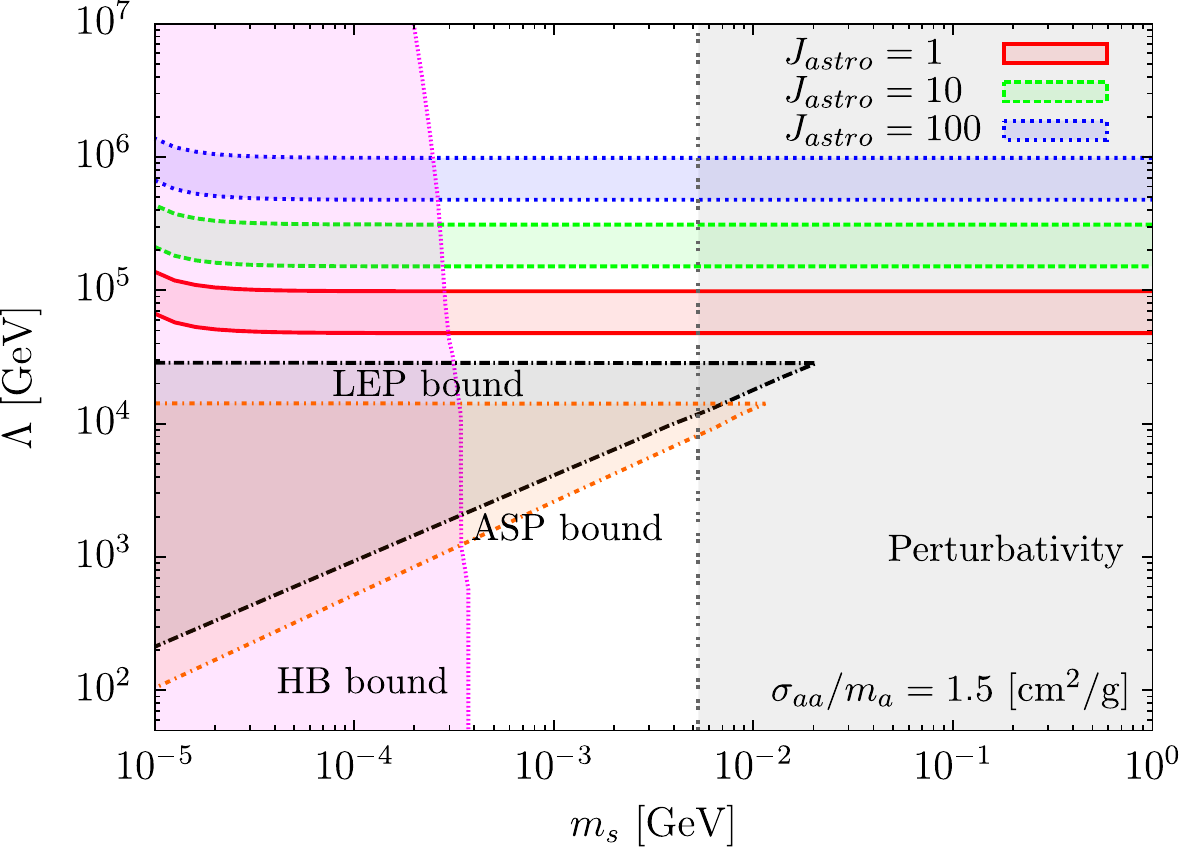}
\caption{Parameter space ($m_s$, $\Lambda$) respecting at the same time
 the 3.5 keV line signal observed by XMM Newton \cite{Bulbulline} 
and two claimed values of self interacting dark matter : 
 $\sigma_{aa}/m_a=1.7\times10^{-4}~\mathrm{cm^2/g}$~\cite{Massey:2015dkw}
 (above), 
 $\sigma_{aa}/m_a=1.5~\mathrm{cm^2/g}$~\cite{Kahlhoefer:2015vua} (below). 
 The factor $J_{astro}$ corresponds to the astrophysical
 parameter. The values of
 $J_{astro}=1~(\mathrm{red}),10~(\mathrm{green}),100~(\mathrm{blue})$
 are taken ($J_{astro}=1$ in the case of an isothermal profile). We also
 represented in the plot the actual limits from the HB star (violet), LEP
 (black), ASP (brown) and the perturbativity (grey).}
\label{Fig:kevline}
\end{center}
\end{figure}

 \noindent
 We show in Fig.~\ref{Fig:idlimit}  and Fig.~\ref{Fig:kevline}
 the combined analysis, including the
 HB stars, ASP and LEP constraints~\cite{Cadamuro:2012rm,Kleban:2005rj}
 for two different values of the
 self-interacting cross-section, $\sigma_{aa}/m_a=1.7 \times
 10^{-4}~\mathrm{cm^2/g}$ and $1.5 ~\mathrm{cm^2/g}$ corresponding to the
 values derived in~\cite{Massey:2015dkw} and~\cite{Kahlhoefer:2015vua} respectively. In
 Fig.~\ref{Fig:idlimit} the analysis is made by taking into account the
 current limits from different observations, 
 whereas in Fig.~\ref{Fig:kevline} we fixed the annihilation
 cross-section to fit the 3.5 keV line
 observation by XMM-Newton Eq.~(\ref{Eq:sigv}).

We would like to insist that our aim is not to affirm that these two
observations are the 
signatures of dark matter, but that combining these two
physical measurements one can deduce a very strong constraint  
and/or information on 
$\Lambda$, especially if one uses the current limits from different
experiments looking at the sky from the keV to the MeV energy range. 
To illustrate our purpose, we can extract lower bounds for the scale of
the Beyond the Standard Model (BSM) $\Lambda$  from the observations of
the satellites HEAO-1, INTEGRAL, COMPTEL, EGRET and FERMI
\cite{Essig:2013goa}.
For our analysis, we required that the photon flux coming from the dark matter
annihilation does not exceed the observed central value 
 plus twice the error bar where the NFW dark matter profile is
 assumed~\cite{Navarro:1996gj}. 
This is depicted in Fig.~\ref{Fig:idlimit} where we
plotted the limit we obtained on $\Lambda$  
 for $\sigma_{aa}/m_a = 1.7
\times 10^{-4} ~ \mathrm{cm^2/g}$ and $\sigma_{aa}/m_a = 1.5~\mathrm{cm^2/g}$
assuming the mass ratio $m_s/m_a=10$. We notice that 
the limits are quite stronger than the ones obtained by LEP,  
especially for large self-interaction cross section.

As one can see from Fig.~\ref{Fig:kevline}, it is interesting to note
that there exists a band of parameter space, 
for $m_s \simeq 1-10$ MeV and $\Lambda \simeq 10-1000$ TeV
where one can  explain the observed 3.5 keV line from
the Perseus cluster for a self-interaction cross section of the order
of magnitude of the one claimed to have been recently observed
and still being largely compatible with accelerator searches.

\subsection{Non-detection of X-ray line}
\begin{figure}
\begin{center}
\includegraphics[scale=0.65]{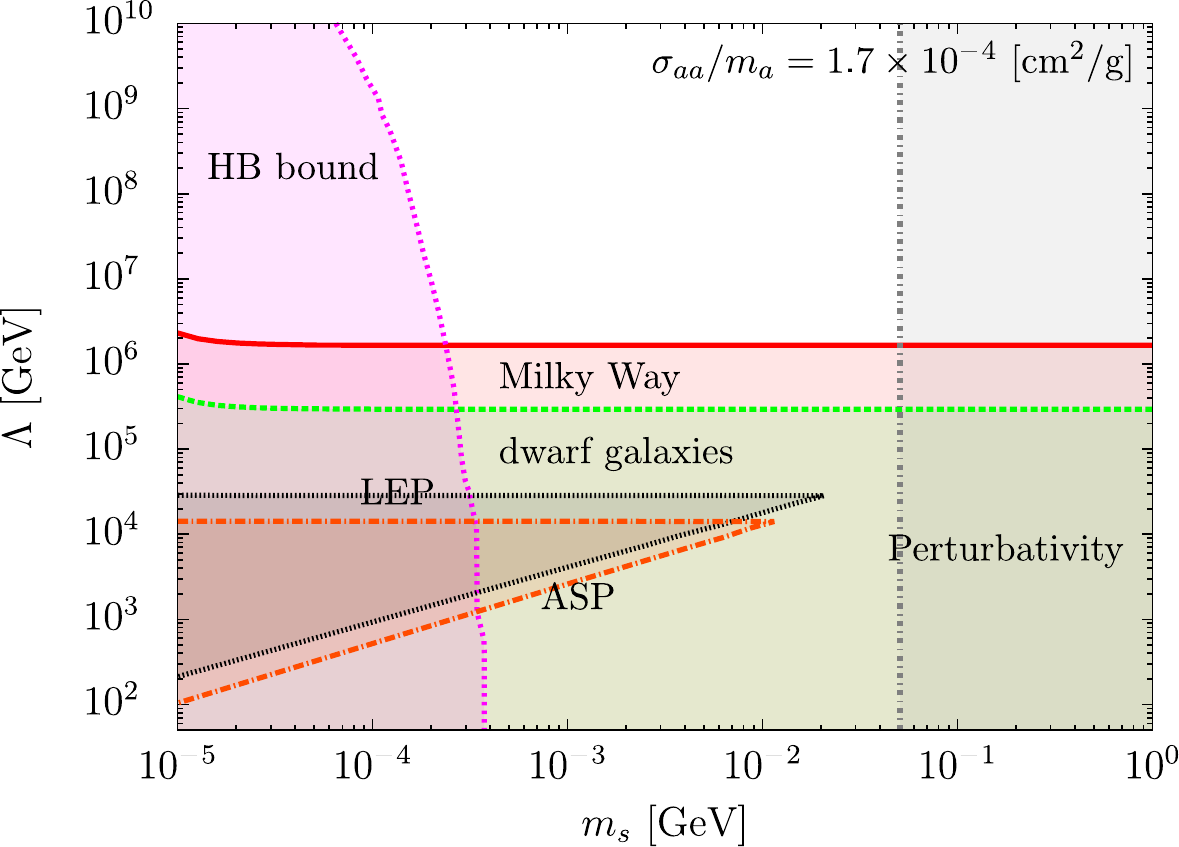}
\includegraphics[scale=0.65]{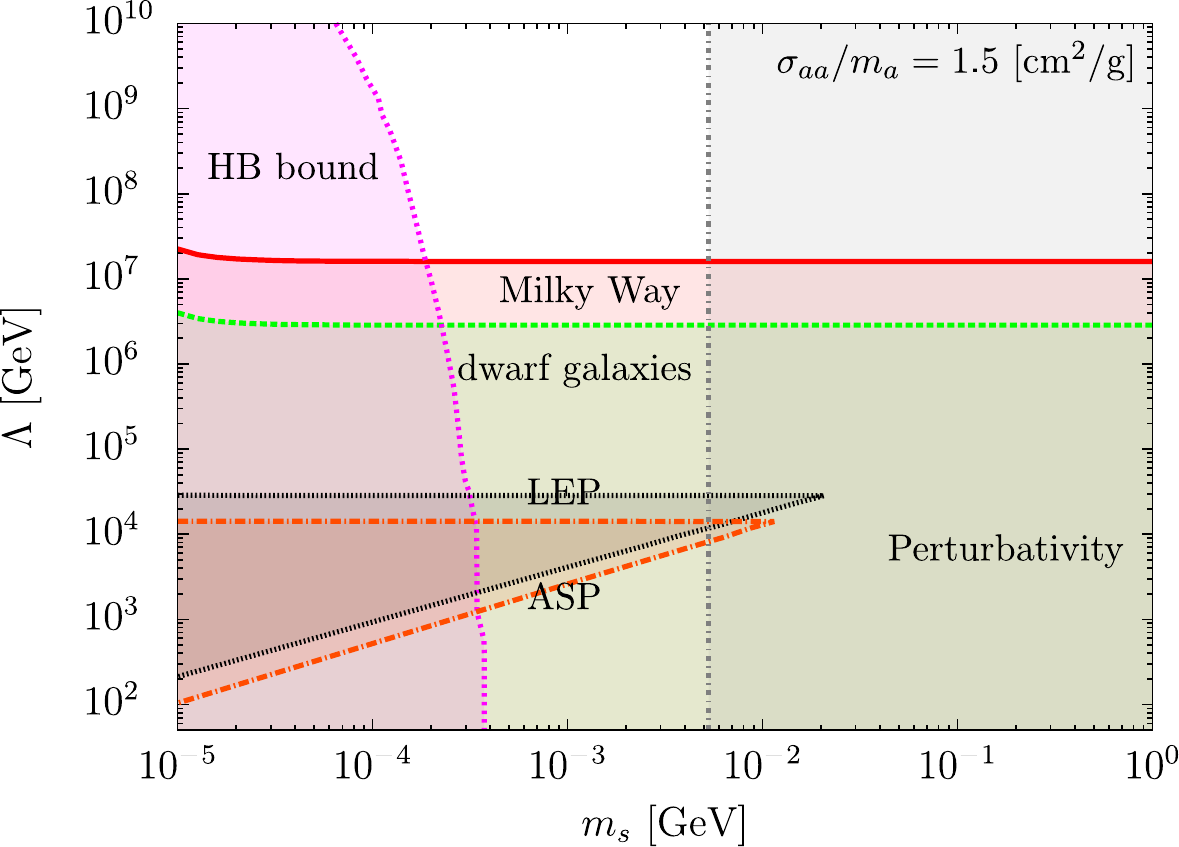}
\caption{Same as Fig.\ref{Fig:kevline} if one considers the 
 non-observation of the 3.5 keV line in the Milky Way 
or dwarf galaxies. 
The region below the red-full line is excluded by the analyses of the
 Milky Way in \cite{Riemer-Sorensen:2014yda} 
whereas the zone below the green-dashed line is excluded by the
 non-observation of the line in stacked dwarf galaxies~\cite{Malyshev:2014xqa}}
\label{Fig:nondetection}
\end{center}
\end{figure}

If the $3.5~\mathrm{keV}$ X-ray line excess discussed above is
interpreted as a dark matter signal, 
a same excess should be observed from the other galaxies such as the Milky
Way, M31 and dwarf spheroidal galaxies in addition to
the Perseus and Centaurus clusters. 
However such an excess has not been observed in the Milky
Way~\cite{Riemer-Sorensen:2014yda}, M31~\cite{Horiuchi:2013noa}, stacked 
galaxies~\cite{Anderson:2014tza} and stacked dwarf galaxies~\cite{Malyshev:2014xqa}. 
For completeness, keeping open all possible interpretation of the 3.5 keV
line signal, we present the result of these analyses in Fig.~\ref{Fig:nondetection}.

This inconsistency can be managed in some models. 
The first example is the scenario of decaying dark matter into axion like
particle~\cite{Conlon:2014wna}. 
In this model, dark matter decays into an axion like particle with 
the energy $3.5~\mathrm{keV}$. 
The axion like particle produced in the process can be converted into photon via
the astrophysical magnetic field around the galaxy clusters. Since the X-ray flux
from dark matter depends on the strength of the magnetic field of each 
galaxy, the non-detection of the X-ray excess in some galaxies can be
consistent. 

Another type of interpretation concerns the possibility of an exciting dark matter ~\cite{Finkbeiner:2014sja}. 
In this scenario, dark matter $\chi$ with the mass of the order of 10
GeV possesses an excited state $\chi^*$. 
The excited state can be produced by up-scattering process
$\chi\chi\to\chi^*\chi^*$ in the center of the cluster and converting the
kinetic energy of dark matter. Then the excited 
state decays into the ground state and photon $\chi^*\to\chi\gamma$. 
One can reproduce the X-ray line excess with the mass difference of
$3.5~\mathrm{keV}$. 
Moreover since the up-scattering process can occur in more massive and
hotter environments such as clusters,
non-detection of the X-ray line excess in smaller galaxies would be
reasonable.

\subsection{Direct detection prospects}

\noindent
Such a light keV-MeV dark matter particle is clearly out of the reach of
any present direct detection technology.
 However recently, the authors of Ref.~\cite{Hochberg:2015pha} proposed a
 new class of superconducting detectors 
 which are sensitive to $\mathcal{O}$(meV) electron recoils from dark
 matter-electron scattering. 
 Such devices could detect dark matter as light as 
 $10$ keV which is exactly the mass 
 range of interest for our model. The idea is to observe the dark matter
 scattering off free electrons in a superconducting metal.
 Indeed, in a superconductor, the free electrons are bound into Cooper
 pairs, which typically have a meV scale (or less) binding energy, which
 is the typical energy transported by a 
 $\mathcal{O}(10)$ keV dark matter with a local
 velocity $\simeq 300~\mathrm{km/s}$. Assuming that $g_e$ is the
 coupling of the electron to the scalar mediator $s$, one 
can straightforwardly compute 
the scattering cross section with an electron $\sigma^e_{\mathrm{DD}}$ :
\begin{equation}
\sigma^e_{\mathrm{DD}}=
\frac{\lambda^2g_e^2}{2\pi
m_s^4}\mu_{ea}^2\left(\frac{m_s^2}{4m_a^2}\right),
\end{equation}
where 
$\mu_{ea}\equiv m_am_e/\left(m_a+m_e\right)$ is the reduced mass.

\begin{figure}
    \begin{center}
    \includegraphics[width=3.in]{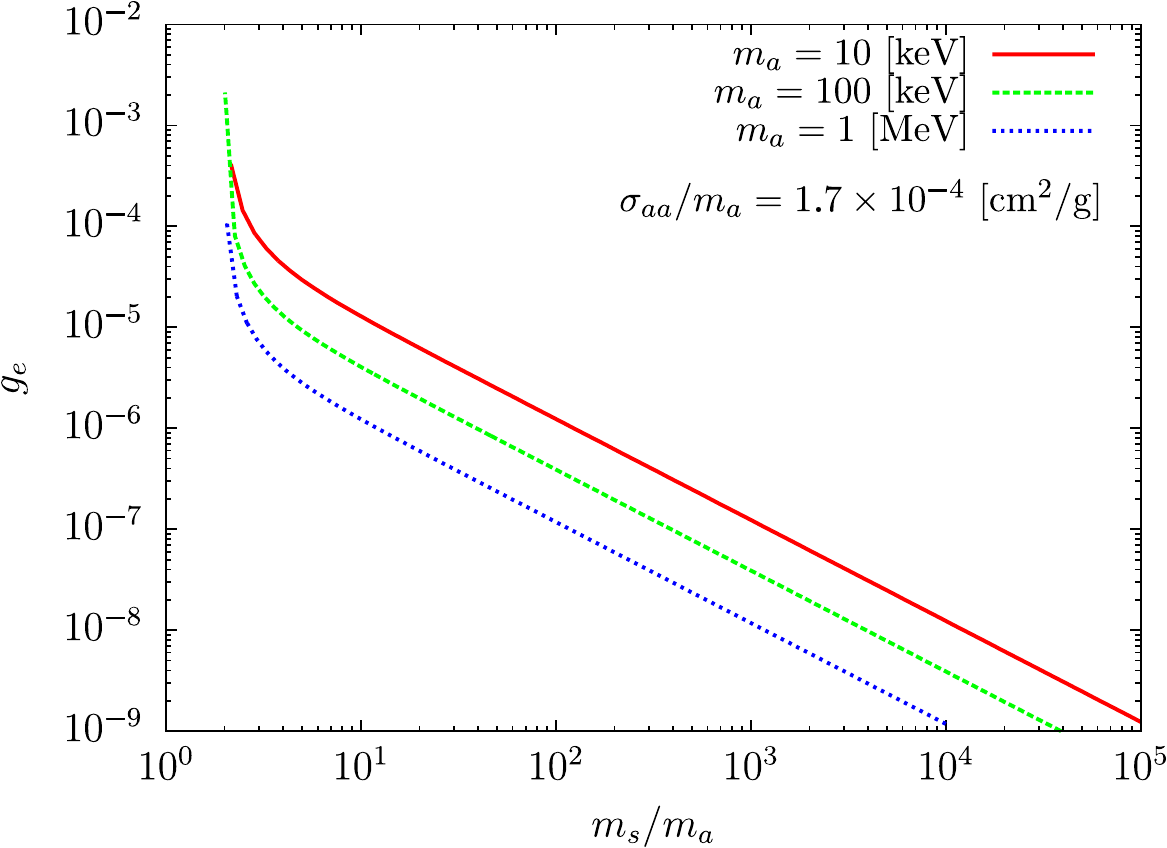}
        \includegraphics[width=3.in]{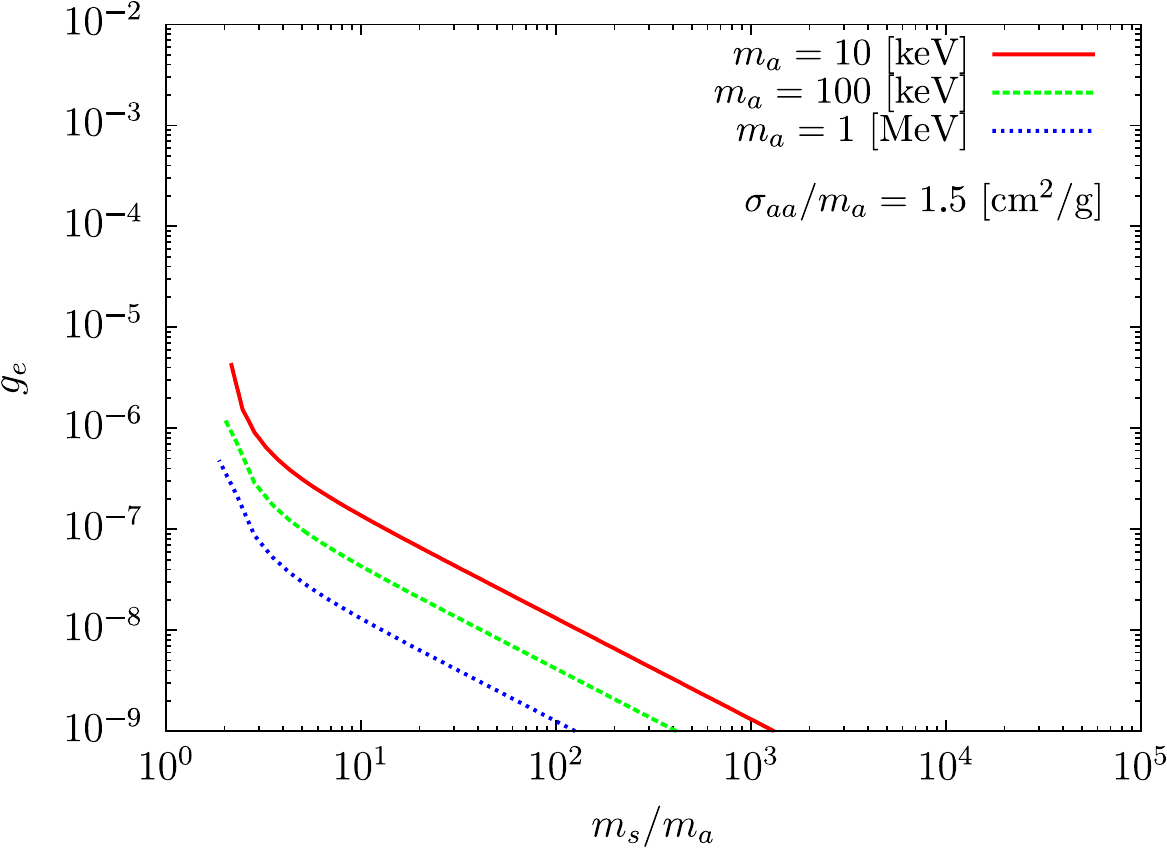}
              \caption{{\footnotesize
              Limits on the electronic coupling to the scalar $s$
     as a function of the ratio $m_s/m_a$ for different values 
              of $m_a$ and a ratio $\sigma_{aa}/m_a=1.7 \times
     10^{-4}~\mathrm{cm^2/g}$ (above) and
     $\sigma_{aa}/m_a=1.5~\mathrm{cm^2/g}$ (below). 
              The curves depict the sensitivity reach of the proposed
     superconducting detectors \cite{Hochberg:2015pha}, 
              for a detector sensitivity to recoil energies between 1
     meV and 1 eV with a kg$\cdot$year of exposure. }}
\label{Fig:ddlimit}
\end{center}
\end{figure}

In Fig.~\ref{Fig:ddlimit}, we show the 95\% expected sensitivity reached
after one
 kg$\cdot$year exposure, corresponding to the cross section required to
 obtain 3.6 signal events~\cite{Feldman:1997qc} supposing a detector
 sensitivity to recoil energies between 1 meV and 1
 eV~\cite{Hochberg:2015pha}. One can see that even for quite low values of 
the coupling $g_e$, the prospect of discovery of self-interacting dark matter through
this new proposal is quite promising.

 \section*{Conclusion}

\noindent
In this work, we have considered a pseudo-scalar dark matter candidate
generated by the breaking of the global $U(1)$ symmetry. 
In this framework, we have shown that one can compellingly combine the
X-ray lines generated by annihilating warm dark matter 
to its self-interacting cross section. 
As a result, we have obtained the limits on the BSM scale $\Lambda\gtrsim
10^5-10^6~\mathrm{GeV}$ and on the dark matter mass
$10~\mathrm{keV}\lesssim m_a\lesssim10~\mathrm{MeV}$ depending on the
fixed self-interacting cross section and the mass ratio $m_s/m_a$. 

Moreover, we have done another combined analysis by fixing the
annihilation cross section in order to reproduce the recent
$3.5~\mathrm{keV}$ line claims. 
Surprisingly, a self-interacting
cross section $\sigma/m$ of the order of 0.1-1 $\mathrm{cm^2/g}$
corresponding to recent claims from the observation of the cluster
Abell 3827
induces naturally a keV line signal corresponding to the one which seems
to have been observed in different clusters
of galaxies like Perseus. Fitting both signals requires a BSM scale of
the order of 100 TeV which could have some consequences
for future accelerator searches.
We have also discussed the non-detection of the X-ray lines from the
Milky Way and stacked dwarf galaxies and found that they give a very
strong constraint on the BSM scale $\Lambda$.

Such a light dark matter can be explored by the recent proposed direct
detection technique through the coupling with electron. 
Even for the small coupling assumed, the detectability of the light dark
matter candidate is promising due to the high sensitivity.

\begin{acknowledgements}
The authors want to thank warmly Giorgio Arcadi, Nicol\'as Bernal,
 Lucien Heurtier and Emilian Dudas for the very 
fruitful discussion during the completion of this work.
This work was also supported by the
Spanish MICINN's Consolider-Ingenio 2010
Programme under grant
Multi-Dark {\bf CSD2009-00064}, the contract {\bf FPA2010-17747},
the France-US PICS no. 06482 and the LIA-TCAP of CNRS. 
Y.~M. acknowledges partial support
 from the European
Union FP7 ITN INVISIBLES (Marie Curie Actions, {\bf PITN-
GA-2011- 289442})
and the ERC advanced grants Higgs@LHC and MassTeV. This research was
 also supported in part by the Research
Executive Agency (REA) of the European Union under
the Grant Agreement {\bf PITN-GA2012-316704} (``HiggsTools'').
The authors would like to thank the Instituto de Fisica Teorica (IFT
 UAM-CSIC) in Madrid for its
support via the Centro de Excelencia Severo Ochoa Program under
Grant {\bf SEV-2012-0249}, during the Program ``Identification of Dark
 Matter with a Cross-Disciplinary Approach'' where some of the ideas presented
in this paper were developed. 
T.~T. acknowledges support from P2IO Excellence Laboratory (Labex).
\end{acknowledgements}

\section*{Appendix}

\noindent
For completeness, here we present the main alternatives to
the pseudo-scalar dark matter candidate, in the same framework.

\subsection*{Majorana dark matter with scalar mediator}
The interaction between a Majorana dark matter and a scalar mediator $s$
is given by
\begin{equation}
\mathcal{L}=-\frac{g_\chi}{2}
 s\overline{\chi^c}\chi+\frac{s}{\Lambda}F_{\mu\nu}F^{\mu\nu},
\end{equation}
with the coupling $g_\chi$.

\noindent
The self-interacting cross section of Majorana dark matter is 
computed from the Yukawa interaction. 
Although there are three contributions to the amplitude coming from $s$,
$t$ and
$u$-channels, the $s$-channel is velocity suppressed and the
other $t$ and $u$-channels give a dominant contribution. 
Thus the self-interacting cross section is given by 

\begin{equation}
\sigma_{\chi\chi}=\frac{g_\chi^4}{8\pi m_\chi^2}\frac{m_\chi^4}{m_s^4},
\end{equation}
where $m_\chi$ is the dark matter mass.

\noindent
For the annihilation process $\chi\chi\to\gamma\gamma$, 
only the $s$-channel contributes and the cross section is given
by~\cite{Dudas:2014ixa}

\begin{equation}
\sigma{v}_{\gamma\gamma}=\frac{g_\chi^2}{\pi\Lambda^2}\frac{m_\chi^4v^2}{(4m_\chi^2-m_s^2)^2}.
\end{equation}

\noindent
As one can see from the formula, this cross section is suppressed by the
dark matter relative velocity $v\sim10^{-3}$. 
Thus it would be difficult to give a connection between the
self-interacting
dark matter and the X-ray monochromatic line unless an enhancement
mechanism like Sommerfeld is taken into account.

\subsection*{Majorana dark matter with pseudo-scalar mediator}

\noindent
For a Majorana dark matter $\chi$ interacting with a pseudo-scalar $a$,
the Lagrangian is given by 

\begin{equation}
\mathcal{L}=-\frac{\tilde{g}_\chi}{2}
 a\overline{\chi^c}\gamma_5\chi+\frac{a}{\Lambda}F_{\mu\nu}\tilde{F}^{\mu\nu}, 
\end{equation}

\noindent
with the coupling $\tilde{g}_\chi$ where $\tilde{F}^{\mu\nu}\equiv
\epsilon^{\mu\nu\rho\sigma}F_{\rho\sigma}/2$ is the dual tensor of
$F_{\mu\nu}$. 
There are $s$, $t$ and $u$-channels contributing to the amplitude for
the self-interacting cross section of dark matter. 
The amplitudes coming from the $t$ and $u$-channels are
velocity-suppressed and negligible. 
As a result, the dark matter self-interacting cross section is given by 

\begin{equation}
\sigma_{\chi\chi}=\frac{\tilde{g}_\chi^4}{8\pi
 m_\chi^2}\frac{m_\chi^4}{(4m_\chi^2-m_a^2)^2}.
\label{eq:self}
\end{equation}

\noindent
The annihilation cross section for $\chi\chi\to\gamma\gamma$ mediated by
the
pseudo-scalar $a$ is computed similarly to the scalar mediator
case~\cite{Dudas:2014ixa}: 

\begin{equation}
\sigma{v}_{\gamma\gamma}=
\frac{4\tilde{g}_\chi^2}{\pi\Lambda^2}\frac{m_\chi^4}{(4m_\chi^2-m_a^2)^2}.
\label{eq:gamma}
\end{equation}

\noindent
From Eq.~(\ref{eq:self}) and (\ref{eq:gamma}), one obtains 

\begin{eqnarray}
\sigma{v}_{\gamma\gamma}&=&
8\sqrt{\frac{2}{\pi}}\frac{m_\chi^{3/2}}{\Lambda^2}
\frac{m_\chi^2}{m_a^2}\sqrt{\frac{\sigma_{\chi\chi}}{m_\chi}}\\
&\simeq&
2.6\times10^{-33}
\left(\frac{10~\mathrm{TeV}}{\Lambda}\right)^2
\left(\frac{E_\gamma}{3~\mathrm{keV}}\right)^{3/2}\nonumber\\
&&\times
\left(\frac{m_\chi/m_a}{0.1}\right)^2
\sqrt{\frac{\sigma_{\chi\chi}/m_\chi}{1~\mathrm{cm^2/g}}}~\mathrm{cm^3/s}. \nonumber
\label{eq:gamma_self}
\end{eqnarray}

\noindent
where $m_\chi\ll m_a$ is assumed. 
One can see that $\Lambda$ is one order of magnitude smaller than the
scalar dark matter case because of the additional suppression due to
the ratio of the squared mass $m_\chi^2/m_a^2$. However, this candidate is also a viable
one, and potentially explain the monochromatic signal and the
self-interacting dark matter in a single framework.


\end{document}